\documentclass{article}
\usepackage{spconf,amsmath,amssymb,amsfonts,bm,mathtools,graphicx}
\usepackage{algorithm,algorithmic}

\def\Dmax{D_{\mathrm{max}}}
\def\Fmat{\mathbf{F}_{\mathcal{T}}}
\def\Sig{\mathbf{\Sigma}}

\graphicspath{{../figures/}}

\title{Dynamic Sensor Pairing for TDOA-Based Target Tracking \\ via mixed-integer Second-Order Cone Programming}

\name{Ryosuke Ikura$^{1}$, Junya Hara$^{1}$, Hiroshi Higashi$^{2}$, Yuichi Tanaka$^{1}$
\thanks{This work is supported in part by JSPS KAKENHI under Grant 26H02536, and JST AdCORP under Grant JPMJKB2307.}}
\address{$^{1}$The University of Osaka, Osaka, Japan~~
  $^{2}$Kansai University, Osaka, Japan}

\begin{document}
% 2段抜きの Table 1 と軌跡の図を同じページの上に置けるようにする（既定は1つ）
\setcounter{dbltopnumber}{2}
\ninept
\maketitle

\begin{abstract}
We propose a sensor pairing method for online target tracking based on time-difference-of-arrival (TDOA) measurements through a sensor network.
Existing pairing designs fix active sensor pairs in advance.
In online tracking, however, adapting the pairing to the current target estimate maintains a favorable sensor-target geometry around it.
We select $K$ pairs at each time step to maximize a Fisher information matrix (FIM) criterion under a communication-degree constraint.
The original combinatorial problem is cast as a mixed-integer second-order cone program.
We solve it at every time step to obtain a globally optimal pairing.
Experiments under various noise settings show that the proposed method obtains the lowest tracking error among the methods under the same communication-degree constraint.
\end{abstract}


\section{Introduction}
\label{sec:intro}
Target tracking estimates the positions of moving targets over time from signals observed by spatially distributed sensors~\cite{li2002detection,brooks2003distributed}.
Its applications include path planning~\cite{chen2004target} and area monitoring~\cite{souza2016target,benfold2011stable}.
One approach to tracking is sequential source localization~\cite{pun2021local,kim2019efficient}, in which the position is estimated at each time instance from the measurements acquired at that instance.
The target is non-cooperative, and the sensors observe when its signal arrives but not when it is transmitted.

We use time-difference-of-arrival (TDOA)~\cite{yaqin2025sensor,liu2015improved}, which measures the differences of the signal arrival times across sensors.
Time-of-arrival~\cite{chakravarty2005multiple,wang2013second} requires the transmission time of the signal to be known, that is, synchronization between the target and the network.
Received signal strength~\cite{feng2011received,xu2014rss} requires calibration of a path-loss model that relates the received power to distance.
In contrast, TDOA requires no synchronization between the target and the network, and therefore is suitable for non-cooperative targets~\cite{so2011source,li2016contributed}.

A TDOA measurement is obtained for a pair of sensors by cross-correlating received waveforms of two sensors.
An $N$-sensor network thus offers $\binom{N}{2}$ candidate measurements, one per pair.
Activating all of them consumes a bandwidth and battery on all coupled sensors~\cite{isler2005sensor,meng2013decentralized,martin2010algorithms}.
The load on sensors is also preferably distributed over the network~\cite{uysal2022sparse}.
In practice, a deployment works under a communication budget, in which $K \ll \binom{N}{2}$ pairs are activated at a time and allows each sensor to communicate with at most $\Dmax$ other sensors.
Choosing the pairs to activate is called \textit{sensor pairing problem}.

Under the communication budget, the accuracy of the position estimate depends on which pairs are activated.
Mathematically, a TDOA constrains the target to one branch of a hyperbola determined by the two sensor positions.
One measurement therefore fixes the position only along the normal to its hyperbola and leaves it undetermined along the tangent.
The $K$ activated pairs define $K$ hyperbolas, and the target lies where they cross.
Since measurement noise biases each hyperbola along its normal, accuracy of the estimate then depends on the sensor-target geometry.
Therefore, the best set of pairs changes as the target moves.

Existing sensor pairing methods fix the pairs in advance~\cite{yaqin2025sensor,khalil2026resourceawaretopologymanagementisacenabled}.
They choose the pairs by optimizing a localization-accuracy criterion.
For example, the pairing design of~\cite{yaqin2025sensor} minimizes the Cramér-Rao bound averaged over the whole surveillance region, and the minimization is carried out by a greedy algorithm. This design seeks good average accuracy over the region at low computational cost.

This design has two limitations for online tracking. 
1) The criterion and the activated pairs are fixed offline and do not depend on the current target position, while adapting the criterion to the current target estimate would maintain a favorable sensor-target geometry.
2) The greedy selection of pairs used in the existing studies is suboptimal and leaves room for improvement.

In this paper, we propose a dynamic sensor pairing method for target tracking.
At every time step, the method selects active pairs by maximizing the determinant of the Fisher information matrix (FIM) under the communication budget.
The FIM is evaluated at the position estimated at the previous time step, under the assumption that the target moves smoothly over time.
The criterion is therefore position-adaptive, and the pairs are updated as the target moves.
Although the selection problem is inherently combinatorial, we can solve it to the global optimum.
For planar tracking, the criterion admits a second-order cone representation, which turns the selection problem into a mixed-integer second-order cone program (MISOCP).
A standard branch-and-bound solver efficiently finds the global optimum with a guarantee of optimality.

Experiments on synthetic sensor networks compare the proposed method with the static design and with a baseline dynamic selection.
Under three noise models, 1) uniform, 2) distance-dependent, and 3) non-line-of-sight noise, the proposed method achieves lower tracking error than existing methods.
We also show that the exact solution is obtained within one second per time step for networks of up to $50$ sensors.

\textit{Notation: } Bold lowercase and uppercase letters denote vectors and matrices, respectively.
We denote the $a$th column and the $(a,b)$ element of a matrix $\mathbf{X}$ by $[\mathbf{X}]_{a}$ and $[\mathbf{X}]_{ab}$, and we use the same bracket notation for the entries of a vector.
Calligraphic letters represent sets of indices, and the superscript $^{\top}$ denotes the transpose.
The $\ell_2$-norm is denoted by $\|\cdot\|$, $\bm{1}$ denotes the all-ones vector, and $\operatorname{diag}(\cdot)$ denotes the diagonal matrix with elements of the input vector on its diagonal, whereas $\operatorname{tr}(\cdot)$, $\det(\cdot)$ and $\nabla$ represent the trace, the determinant and the gradient, respectively.

\section{Preliminaries}
\label{sec:pre}
We consider a two-dimensional setting with a single target at the true position $\bm{p}\in\mathbb{R}^2$, and we write the $i$th sensor position as $\bm{s}_i\in\mathbb{R}^2~(i=1,\ldots,N)$.
For a sensor pair $(i,j)$, the noiseless time difference is
\begin{equation} \label{eq:tau}
    \tau_{ij}(\bm{p}) = (\|\bm{p}-\bm{s}_i\| - \|\bm{p}-\bm{s}_j\|)/c,
\end{equation}
where $c$ is the propagation speed~\cite{yaqin2025sensor}.
Without loss of generality, we set $c=1$.

In this paper, we assume that $\tau_{ij}$ is observed with noise as $z_{ij}=\tau_{ij}(\bm{p})+n_{ij}$, where $n_{ij}\sim\mathcal{N}\big(0,\sigma_{ij}^2(\bm{p})\big)$ is independent across pairs.
The variance $\sigma_{ij}^2(\bm{p})$ grows with the target-to-sensor distances~\cite{torrieri2007statistical,zhao2019sensor}, i.e.,
\begin{equation} \label{eq:noise}
    \sigma_{ij}^2(\bm{p})=\kappa\big(\|\bm{p}-\bm{s}_i\|^\eta+\|\bm{p}-\bm{s}_j\|^\eta\big),
\end{equation}
where $\kappa>0$ sets the noise level and $\eta\ge 0$ selects the propagation regime: $\eta=0$ gives a uniform model and $\eta=2$ a noise that depend on the Euclidean distance between the target and the sensors~\cite{zhao2019sensor}.

Let $\mathcal{T}$ be the set of active pairs, and let vectors $\bm{\tau}_{\mathcal{T}}(\bm{p})$, $\bm{z}_{\mathcal{T}}$, and $\bm{\sigma}^2_{\mathcal{T}}(\bm{p})$ collect $\tau_{ij}$, $z_{ij}$, and $\sigma_{ij}^2$ for all pairs in $\mathcal{T}$.
Since the noise is independent across pairs, the covariance of $\bm{z}_{\mathcal{T}}$ is $\Sig_{\mathcal{T}}(\bm{p})=\operatorname{diag}\big(\bm{\sigma}^2_{\mathcal{T}}(\bm{p})\big)$, where $\sigma_k^2=\sigma_{ij}^2$ is the entry of $\bm{\sigma}^2_{\mathcal{T}}$ for the pair $k$ coupling sensors $i$ and $j$.
The measurements are therefore distributed as $\bm{z}_{\mathcal{T}}\sim\mathcal{N}\big(\bm{\tau}_{\mathcal{T}}(\bm{p}),\Sig_{\mathcal{T}}(\bm{p})\big)$.
% and the target position is estimated by maximizing their likelihood~\cite{ma2021maximum,chan1994simple}, that is,
% \begin{equation} \label{eq:ml}
% \begin{split}
%     \hat{\bm{p}} = \underset{\bm{p}}{\operatorname{argmin}}~&\big(\bm{z}_{\mathcal{T}}-\bm{\tau}_{\mathcal{T}}(\bm{p})\big)^\top\Sig_{\mathcal{T}}(\bm{p})^{-1}\big(\bm{z}_{\mathcal{T}}-\bm{\tau}_{\mathcal{T}}(\bm{p})\big)\\
%     &+\log\det\big(\Sig_{\mathcal{T}}(\bm{p})\big).
% \end{split}
% \end{equation}
% The accuracy attainable in~\eqref{eq:ml} is characterized by the FIM of $\bm{p}$.
Since both the mean and the covariance depend on $\bm{p}$, the FIM is given by the Slepian--Bangs formula~\cite{kay1993statistical,torrieri2007statistical}.
Let $\mathbf{J}_{\mathcal{T}}$ be the Jacobian of $\bm{\tau}_{\mathcal{T}}$ with respect to $\bm{p}$, where $[\mathbf{J}_{\mathcal{T}}]_k = \bm{w}_k^{\top}=(\bm{u}_i-\bm{u}_j)^{\top}$ with the unit bearing vector $\bm{u}_i=(\bm{p}-\bm{s}_i)/\|\bm{p}-\bm{s}_i\|$.
Then the $(a,b)$ entry of $2 \times 2$ matrix $\Fmat(\bm{p})$ is given by
\begin{equation} \label{eq:fim_full}
    [\Fmat(\bm{p})]_{ab} = [\mathbf{J}_{\mathcal{T}}]_{a}^{\top}\Sig_{\mathcal{T}}^{-1}[\mathbf{J}_{\mathcal{T}}]_{b}
    +\tfrac{1}{2}\operatorname{tr}\!\left(\Sig_{\mathcal{T}}^{-1}\Sig_{\mathcal{T},a}\Sig_{\mathcal{T}}^{-1}\Sig_{\mathcal{T},b}\right),
\end{equation}
where $\Sig_{\mathcal{T},a}=\partial\Sig_{\mathcal{T}}/\partial[\bm{p}]_a$, and every quantity on the right-hand side is evaluated at $\bm{p}$.
The first term of~\eqref{eq:fim_full} corresponds to the sensitivity of the measurement mean to $\bm{p}$.
The second term corresponds to the sensitivity of the covariance, which depends on $\bm{p}$ through~\eqref{eq:noise}.
The Cram\'er-Rao lower bound at the true position is $\Fmat^{-1}(\bm{p})$~\cite{vantrees2001detection}, and hence maximizing an optimality criterion of $\det(\Fmat)$ tightens the bound on the estimation error.

\section{Dynamic Sensor Pairing for Target Tracking}
\label{sec:proposed}
In this section, we present the proposed dynamic sensor pairing method.
We formulate a pairing problem solved at each time step, show that this combinatorial problem is written exactly as a mixed-integer second-order cone program (MISOCP), and apply the resulting pairing to online tracking.

\begin{figure}[t]
    \centering
    \includegraphics[width=\linewidth]{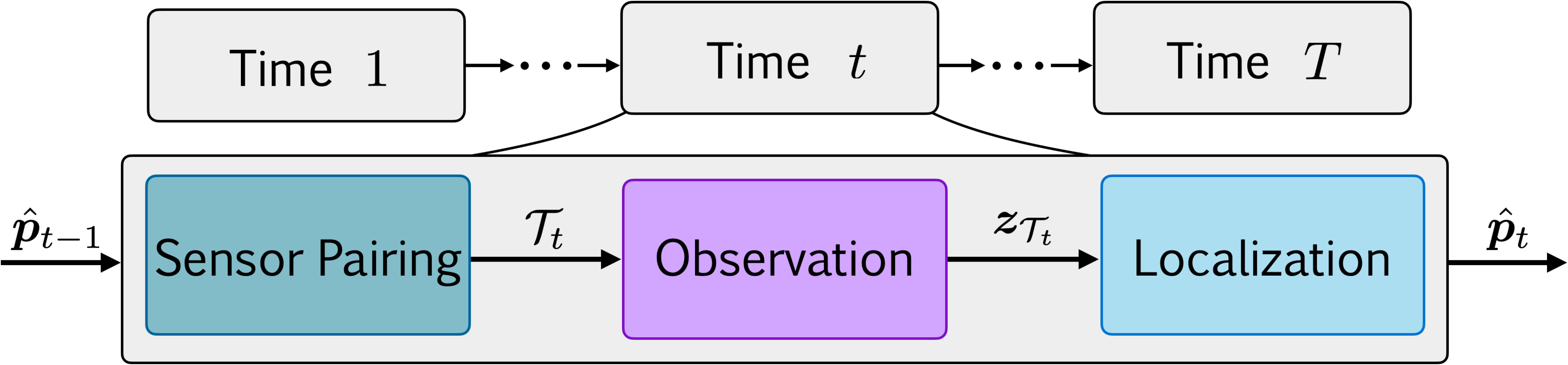}
    \caption{Framework for online target tracking, where $T$ is the total number of time steps and the lower panel shows the processing at a single time step $t$.}
    \label{fig:framework}
\end{figure}

We first describe the framework for online target tracking, illustrated in Fig.~\ref{fig:framework}.
Let $t$ denote the time index, and let $\hat{\bm{p}}_t\in\mathbb{R}^2$ be the target position estimated at time $t$.
At the beginning of time step $t$, the active sensor pairs $\mathcal{T}_t$ are determined.
The system activates only those pairs and collects the corresponding TDOA measurements as $\bm{z}_{\mathcal{T}_t}$.
The localization step then updates $\hat{\bm{p}}_{t-1}$ to $\hat{\bm{p}}_t$ using $\bm{z}_{\mathcal{T}_t}$.
This procedure is repeated over time, so that the sensor pairing is dynamically updated.

\subsection{Formulation of Sensor Pairing Problem}
We now formulate the sensor pair selection problem at time step $t$.
We assume that the target moves smoothly between consecutive time steps.
Therefore, the FIM evaluated at $\hat{\bm{p}}_{t-1}$ is a good approximation of the FIM at the true position $\bm{p}_t$.

Sensor networks are represented as unweighted graphs, where sensors and sensor pairs correspond to nodes and edges, respectively.
Let $\mathcal{V}$ be the set of $N$ sensors and $\mathcal{E}$ the set of all $\binom{N}{2}$ candidate pairs, indexed by $k$.
Pairing then specifies a subset $\mathcal{T}\subset\mathcal{E}$ with $|\mathcal{T}|=K$.

The FIM of a given pairing $\mathcal{T}$ decomposes into the sum of per-pair contributions.
Since $\Sig_{\mathcal{T}}$ is diagonal, so are $\Sig_{\mathcal{T}}^{-1}$ and $\Sig_{\mathcal{T},a}=\operatorname{diag}\big(\partial\bm{\sigma}^2_{\mathcal{T}}/\partial[\bm{p}]_a\big)$, whose diagonal entries for the pair $k$ are $1/\sigma_k^2$ and $[\bm{g}_k]_a$, where $\bm{g}_k\triangleq\nabla_{\bm{p}}\sigma_k^2$, respectively.
Substituting $\Sig_{\mathcal{T}}^{-1}$ and $\Sig_{\mathcal{T},a}$ into~\eqref{eq:fim_full}, we obtain 
\begin{equation} \label{fim_entry}
    [\Fmat(\bm{p})]_{ab}=\sum_{k\in\mathcal{T}}\big(\frac{[\bm{w}_k]_a[\bm{w}_k]_b}{\sigma_k^2}+\frac{[\bm{g}_k]_a[\bm{g}_k]_b}{2\sigma_k^4}\big),
\end{equation}
for every index pair $(a,b)$.
Collecting the entries into a matrix gives $\Fmat(\hat{\bm{p}}_{t-1})=\sum_{k\in\mathcal{T}}\mathbf{B}_k$ with $\mathbf{B}_k=\bm{w}_k\bm{w}_k^\top/\sigma_k^2+\bm{g}_k\bm{g}_k^\top/2\sigma_k^4$, where $\bm{w}_k$, $\sigma_k^2$, and $\bm{g}_k$ are evaluated at the previous position estimate $\hat{\bm{p}}_{t-1}$.

Following~\cite{yaqin2025sensor}, we adopt the D-optimality criterion $\det(\Fmat)$ for sensor pairing.
The pairing problem at time step $t$ is thus given by
\begin{equation} \label{eq:problem}
\begin{split}
    &\mathcal{T}^* = \underset{\mathcal{T}\subset\mathcal{E}}{\operatorname{argmax} ~}\det\big(\Fmat(\hat{\bm{p}}_{t-1})\big)\\
    &\text{s.t.}|\mathcal{T}|=K,~\max_i\,[\bm{d}(\mathcal{T})]_i\le \Dmax,    
\end{split}
\end{equation}
where $[\bm{d}(\mathcal{T})]_i$ is the degree of node $i$ with respect to the edge set $\mathcal{T}$.
In practice, each sensor communicates with a limited number of other sensors because of power and bandwidth limitations~\cite{yaqin2025sensor}.
We therefore impose a maximum degree $\Dmax$ on the sensors.
This problem is combinatorial, with at most $\binom{|\mathcal{E}|}{K}$ candidate subsets.

\subsection{Problem Reformulation}
\label{subsec:conic}
While~\eqref{eq:problem} should be solved exactly, it is a combinatorial problem and the brute-force search is impractical for an online implementation.
We reformulate~\eqref{eq:problem} into a MISOCP~\cite{sagnol2015computing} in three steps: 1) indicator vector formulation, 2) three-scalar parameterization, and 3) cone representation.

\noindent\textbf{Step 1: Indicator Vector Formulation}

We rewrite~\eqref{eq:problem} using an indicator vector $\bm{m}\in\{0,1\}^{|\mathcal{E}|}$, where $[\bm{m}]_k=1$ if $k\in\mathcal{T}$ and $0$ otherwise.
Since the FIM is expressed as a summation of per-pair contributions, the pairing problem becomes
\begin{equation} \label{eq:integer}
    \bm{m}^* = \underset{\bm{m}\in\{0,1\}^{|\mathcal{E}|}}{\operatorname{argmax}}\det\big(\mathbf{F}(\bm{m})\big)
    \text{,\quad s.t. }\bm{1}^\top\bm{m}=K,~\mathbf{\Phi}\bm{m}\le \Dmax\bm{1},
\end{equation}
where $\mathbf{F}(\bm{m})=\sum_{k=1}^{|\mathcal{E}|}[\bm{m}]_k\mathbf{B}_k$ is the FIM in~\eqref{eq:problem} written as a function of the selection, and $\mathbf{\Phi}\in\{0,1\}^{|\mathcal{V}|\times|\mathcal{E}|}$ is the node-edge incidence matrix, i.e., $[\mathbf{\Phi}]_{pq}=1$ if node $p$ is an endpoint of edge $q$ and $0$ otherwise.
Accordingly, $\mathbf{\Phi}\bm{m}=\bm{d}(\mathcal{T})$.

\noindent\textbf{Step 2: Three-Scalar Parameterization}

We rewrite $\det\bigl(\mathbf{F}(\bm{m})\bigr)$ as a quadratic form in three scalars, each of which is linear in $\bm{m}$, and $\bm{m}$
is subject to a second-order cone constraint.
Since a $2\times 2$ symmetric matrix $\mathbf{X}$ has three independent entries, we set
\begin{equation} \label{eq:coords}
    f_0(\mathbf{X}) \triangleq \tfrac{[\mathbf{X}]_{11}+[\mathbf{X}]_{22}}{2},\,
    f_1(\mathbf{X}) \triangleq \tfrac{[\mathbf{X}]_{11}-[\mathbf{X}]_{22}}{2},\,
    f_2(\mathbf{X}) \triangleq [\mathbf{X}]_{12}.
\end{equation}
A direct calculation of $\text{det}(\mathbf{F}(\bm{m}))$ gives:
\begin{equation} \label{eq:det}
    \det\big(\mathbf{F}(\bm{m})\big) = f_0(\mathbf{F}(\bm{m}))^2-f_1(\mathbf{F}(\bm{m}))^2-f_2(\mathbf{F}(\bm{m}))^2 .
\end{equation}
Thus $\mathbf{F}(\bm{m})$ is positive semidefinite if and only if
\begin{equation} \label{eq:psd}
    \sqrt{f_1(\mathbf{F}(\bm{m}))^2+f_2(\mathbf{F}(\bm{m}))^2}\le f_0(\mathbf{F}(\bm{m})).
\end{equation}
Since all $f_\ell$ in~\eqref{eq:coords} are linear, the linearity of $\mathbf{F}(\bm{m})$ carries over to the three scalars, where $f_\ell(\mathbf{F}(\bm{m}))=\sum_{k=1}^{|\mathcal{E}|}f_\ell(\mathbf{B}_k)[\bm{m}]_k$ for $\ell=0,1,2$.
The objective of~\eqref{eq:integer} is thus the quadratic form~\eqref{eq:det} in three linear functionals of $\bm{m}$.

\noindent\textbf{Step 3: Cone Representation}

We introduce an auxiliary variable $\gamma\ge 0$ and rewrite the determinant objective through the hypograph of $\sqrt{\det\big(\mathbf{F}(\bm{m})\big)}$.
Since every $\mathbf{B}_k$ is positive semidefinite, $f_0(\mathbf{F}(\bm{m}))=\tfrac{1}{2}\operatorname{tr}\mathbf{F}(\bm{m})\ge 0$.
With this, \eqref{eq:det} shows that the hypograph, namely the set of $(\bm{m},\gamma)$ with $\gamma^2\le\det\big(\mathbf{F}(\bm{m})\big)$, is described by
\begin{equation} \label{eq:cone}
    \sqrt{f_1(\mathbf{F}(\bm{m}))^2+f_2(\mathbf{F}(\bm{m}))^2+\gamma^2}\;\le\; f_0(\mathbf{F}(\bm{m})) ,
\end{equation}
which is a second-order cone constraint in $\bm{m}$ and $\gamma$~\cite{lobo1998applications}.
Maximizing $\gamma$ under~\eqref{eq:cone} maximizes $\det\big(\mathbf{F}(\bm{m})\big)$ since the square root function is increasing.
Finally, collecting the constraints, the pairing problem~\eqref{eq:integer} becomes
\begin{equation} \label{eq:misocp}
\begin{split}
    \underset{\bm{m},\gamma}{\text{max}}&\quad  \gamma \\
    \text{s.t.}&\quad  \sqrt{f_1(\mathbf{F}(\bm{m}))^2+f_2(\mathbf{F}(\bm{m}))^2+\gamma^2}\le f_0(\mathbf{F}(\bm{m})),\\
    &\bm{1}^\top\bm{m}=K,\quad \mathbf{\Phi}\bm{m}\le \Dmax\bm{1},\quad \bm{m}\in\{0,1\}^{|\mathcal{E}|},\quad \gamma\ge 0 .
\end{split}
\end{equation}
With a linear objective and binary selection variables, \eqref{eq:misocp} is a MISOCP, and a branch-and-bound solver~\cite{land1960automatic} returns a global optimum of~\eqref{eq:problem} whenever a feasible pairing exists.

\subsection{Application to Online Tracking}
\label{subsec:application}

\begin{algorithm}[t]
    \caption{Online target tracking with dynamic sensor pairing}
    \label{alg:tracking}
    \begin{algorithmic}[1]
    \REQUIRE Initial estimate $\hat{\bm{p}}_{0}$, sensor positions $\{\bm{s}_i\}$, number of pairs $K$, maximum degree $\Dmax$.
    \FOR{$t=1,2,\ldots$}
        \STATE Evaluate the FIM at $\hat{\bm{p}}_{t-1}$ in~\eqref{eq:integer} and form the cone coefficients $f_\ell(\mathbf{B}_k)$.
        \STATE Obtain $\mathcal{T}_{t}$ by solving the MISOCP~\eqref{eq:misocp}.
        \STATE Activate the pairs in $\mathcal{T}_{t}$ and acquire $\bm{z}_{\mathcal{T}_t}$.
        \STATE Estimate the target position from $\bm{z}_{\mathcal{T}_t}$ and update $\hat{\bm{p}}_{t}$.
    \ENDFOR
    \end{algorithmic}
\end{algorithm}

Solving~\eqref{eq:misocp} at time step $t$ determines the active pairs.
Algorithm~\ref{alg:tracking} summarizes the resulting framework for online tracking.
The initial estimate $\hat{\bm{p}}_0$ can be arbitrary.

\begin{table}[t]
\centering
\caption{Tracking root-mean-square error (RMSE) (mean $\pm$ standard deviation over $50$ trials, lower is better) for $N=10$ sensors, $K=9$ active pairs and maximum degree $\Dmax=5$, under three noise models. Bold marks the best degree-constrained method in each column.}
\label{tab:comparison}
\footnotesize
\begin{tabular}{lccc}
\hline
Method & Uniform & Distance &  NLOS \\
\hline
Proposed & \textbf{0.09 $\pm$ 0.05} & \textbf{0.14 $\pm$ 0.11} & \textbf{0.51 $\pm$ 0.22} \\
Static pairing~\cite{yaqin2025sensor} & 0.11 $\pm$ 0.07 & 0.22 $\pm$ 0.16 & 0.55 $\pm$ 0.25 \\
NES & 0.17 $\pm$ 0.18 & 0.19 $\pm$ 0.18 & 0.67 $\pm$ 0.37 \\
Random & 0.15 $\pm$ 0.08 & 0.24 $\pm$ 0.16 & 0.56 $\pm$ 0.24 \\
\hline
All pairs$^{\dagger}$ & 0.06 $\pm$ 0.03 & 0.09 $\pm$ 0.06 & 0.37 $\pm$ 0.15 \\
\hline
\end{tabular}
\\[2pt]
\footnotesize $^{\dagger}$ Shown as an unconstrained reference.
\end{table}

\begin{figure}[t]
    \centering
    \includegraphics[width=\linewidth]{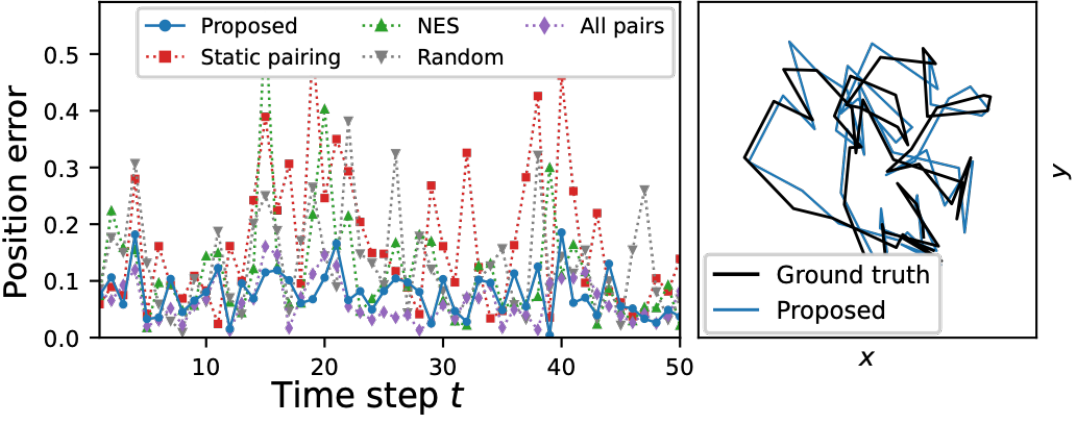}
    \caption{Position error at each time step (left) and the true target trajectory with the trajectory estimated by the proposed method, enlarged around the path (right), in one trial under the distance-dependent noise model, with $N=10$, $K=9$, and $\Dmax=5$.}
    \label{fig:trajectory}
\end{figure}

\section{Experiments}
\label{sec:exp}
In this section, we perform target tracking simulations with synthetic sensor networks.
These experiments are conducted based on Algorithm~\ref{alg:tracking}, and we evaluate the tracking accuracy, the effect of the communication budget and the pairing time.

\subsection{Setting}
We consider a two-dimensional square region $[0,10]^2$, where $N=10$ sensors are randomly placed.
The target motion is generated so that current position $\bm{p}_t$ is close to previous one $\bm{p}_{t-1}$.
Specifically, $\bm{p}_t$ is sampled within a disk of radius $0.5$ centered at $\bm{p}_{t-1}$ each time, i.e., the target randomly wanders.

At time step $t$, we activate $K$ sensor pairs and acquire the corresponding TDOA measurements.
Following~\cite{yaqin2025sensor}, we set $K=N-1=9$.
In addition, we set $\Dmax=5$.
As a position estimation algorithm, we employ a Gauss--Newton iteration based on Taylor-series linearization~\cite{foy1976position}.
We set $\hat{\bm{p}}_0$ to the center of the region.

We test three noise models on the TDOA measurements:
\begin{enumerate}\setlength{\itemsep}{0pt}\setlength{\parskip}{0pt}\setlength{\topsep}{2pt}
    \item \textbf{Uniform noise model:} All pairs share the same noise variance, which corresponds to $\eta=0$ in~\eqref{eq:noise}.
    We set $\kappa=10^{-2}$.
    This setting is an ideal environment in which hardware-induced noise is dominant.
    \item \textbf{Distance-dependent noise model:} The noise variance grows with the target-to-sensor distances as in~\eqref{eq:noise} with $\eta=2$.
    We set $\kappa=10^{-3}$.
    This setting reflects realistic propagation effects of sinals.
    \item \textbf{Distance-dependent noise model with non-line-of-sight (NLOS) propagation:} Each sensor is independently obstructed with probability $0.2$ at time step $t$.
    An obstructed sensor $i$ receives the signal through reflected paths, which adds a range bias $b_i\ge 0$ with density $2e^{-2b_i}$ and scales its variance by $\alpha_i=4$.
    The observation of pair $(i,j)$ is then
    \begin{equation} \label{eq:nlos}
    \begin{split}
        z_{ij} &= (\|\bm{p}-\bm{s}_i\|+b_i)-(\|\bm{p}-\bm{s}_j\|+b_j)+n_{ij},\\
        n_{ij} &\sim \mathcal{N}\big(0,\kappa(\alpha_i\|\bm{p}-\bm{s}_i\|^\eta+\alpha_j\|\bm{p}-\bm{s}_j\|^\eta)\big).
    \end{split}
    \end{equation}
    We set $\eta=2$ and $\kappa=10^{-3}$ as in the second model.
    Obstruction is unknown to the system, and hence both the pairing and the estimator use the line-of-sight variance~\eqref{eq:noise}.
    This setting therefore tests robustness to a misspecified noise model.
\end{enumerate}

The proposed method is compared with four baselines:
\begin{enumerate}\setlength{\itemsep}{0pt}\setlength{\parskip}{0pt}\setlength{\topsep}{2pt}
    \item \textbf{Static sensor pairing~\cite{yaqin2025sensor}:} This method selects one pairing independently of the target position, by minimizing the Cram\'er--Rao bound averaged over the surveillance region, and we keep the selected pairing unchanged for the whole track.
    \item \textbf{Nearest-edge selection (NES):} This rule-based dynamic strategy selects pairs in ascending order of the distance $\|\hat{\bm{p}}_{t-1}-\bm{s}_i\|+\|\hat{\bm{p}}_{t-1}-\bm{s}_j\|$ to the previous estimate, provided that neither sensor exceeds the maximum degree $\Dmax$.
    \item \textbf{Random selection:} This method draws $K$ pairs uniformly at random under the same cardinality and degree constraints, and therefore uses no geometric information.
    \item \textbf{All-pairs activation:} This method activates all $\binom{N}{2}$ pairs, ignoring the maximum degree constraint, and serves as an unconstrained reference.
\end{enumerate}

Within a trial, all methods share the same sensor layout, target trajectory, and measurements.
We perform target tracking for $50$ time steps and compute the RMSE of the estimated position.
This process is repeated over $50$ independent trials, and the performance is evaluated by averaging the RMSE over the trials.
We also measure the average running time of the pairing step for each method.
We solve~\eqref{eq:misocp} with SCIP~\cite{bolusani2024scip} through CVXPY~\cite{diamond2016cvxpy}, and all timings are measured on an Apple MacBook Pro with an M2 Pro chip.

\subsection{Results}
\label{subsec:accuracy}
Table~\ref{tab:comparison} shows the tracking RMSE under the three noise models.
We observe that the proposed method achieves the lowest RMSE among the degree-constrained methods in all three cases.
These results show that selecting the pairs by the FIM criterion at the current estimate lowers the tracking error under the same communication budget.
Under the same noise models, the proposed method reaches an RMSE within a factor of $1.6$ of the all-pairs reference while activating a fifth of the candidate pairs.

We also observe that NES leads the static pairing~\cite{yaqin2025sensor} under distance-dependent noise and trails it under uniform noise.
This is likely because choosing sensors close to the target reduces the effect of noise only when the noise variance increases with distance.
Under the NLOS model, the bias of obstructed links violates the zero-mean Gaussian assumption behind the FIM.
The proposed method still obtains the lowest RMSE among the degree-constrained methods, which suggests that our selection is effective even under a misspecified variance model.

Fig.~\ref{fig:trajectory} shows the position error at each time step in one trial under the distance-dependent noise model.
We observe that the error of the proposed method stays small throughout the trajectory, whereas the other degree-constrained methods sometimes produce large errors.

\begin{figure}[t]
    \centering
    \includegraphics[width=\columnwidth]{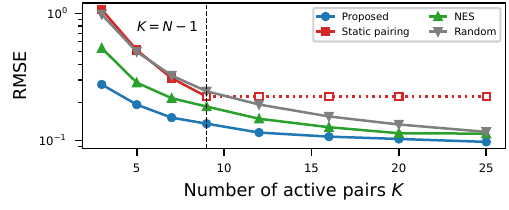}
    \caption{Tracking RMSE against the number of active pairs $K$ under the distance-dependent noise model, with $N=10$ and $\Dmax=5$.
    The dashed line marks $K=N-1$, and the dotted segment beyond it repeats the static design value.}
    \label{fig:cardinality}
\end{figure}

Fig.~\ref{fig:cardinality} shows the tracking RMSE against the number of active pairs $K$ under the distance-dependent noise model.
We observe that the proposed method is better than the other ones, and that its gap to the static design widens as $K$ decreases.
These results indicate that the error relies on the choice of pairs when only a few of them may be activated.

\begin{figure}[t]
    \centering
    \includegraphics[width=\columnwidth]{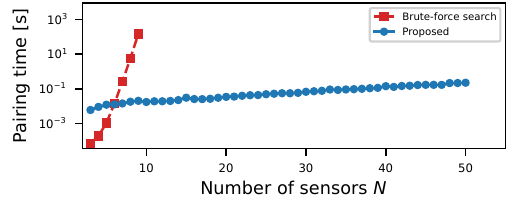}
    \caption{Pairing time per time step against the number of sensors $N$, with $K=N-1$ and $\Dmax=5$, on a logarithmic vertical axis.}
    \label{fig:runtime}
\end{figure}

Fig.~\ref{fig:runtime} compares the pairing time of the proposed method with that of the brute-force search for growing $N$.
We observe that the pairing time of the brute-force search increases rapidly with $N$, whereas that of the proposed method increases slowly.
Both methods obtain the same optimal value in all measured instances.
Moreover, up to $N=50$, the measured pairing time for the proposed method stays below $300$~ms on average and below $1000$~ms for the worst-case instance.
This can be used in practice because the TDOA positioning system typically operates with periods of $450$ to $1000$~ms~\cite{chen2019acoustic}.

\section{Conclusion}
\label{sec:conclusion}
We proposed a dynamic sensor pairing method for TDOA-based online target tracking under communication constraints.
Under the assumption that the target moves smoothly over time, we maximize the information that the active pairs carry about the neighborhood of the previous position estimate.
The pairs at time step $t$ are determined by maximizing the D-optimality criterion of the FIM under a cardinality constraint and a maximum degree constraint.
We showed that this problem is exactly a MISOCP, and hence a globally optimal pairing is obtained within every tracking step.
Simulation results showed that the proposed method obtains the lowest tracking error among degree-constrained methods under three noise models, with the optimum obtained within one second per time step for networks of up to $50$ sensors.

% 2段組では \pagebreak は段を折るだけなので、参考文献を新ページに送るには
% \clearpage を使う（未処理のフロートも同時に掃き出される）。
\clearpage

\bibliographystyle{IEEEbib}
\bibliography{TDOA}

\end{document}